%% file: main.tex
\documentclass[screen]{acmart}
\AtBeginDocument{%
  \providecommand\BibTeX{{%
    \normalfont B\kern-0.5em{\scshape i\kern-0.25em b}\kern-0.8em\TeX}}}

\usepackage{multirow}
\usepackage{tabularx}
\usepackage{makecell}
\usepackage{gensymb}
\usepackage{colortbl}
\usepackage{enumitem}
\usepackage{wrapfig}
\newcommand{\systemname}{InfoPrint}
\newcommand{\ie}{\textit{i.e., }}
\newcommand{\eg}{\textit{e.g., }}
\newcommand{\etal}{\textit{et al. }}

\setcopyright{none}
\acmBooktitle{Submission to ACM}
\begin{document}

%%
%% The "title" command has an optional parameter,
%% allowing the author to define a "short title" to be used in page headers.
\title{\systemname: Embedding Information into 3D Printed Objects}

%%
%% The "author" command and its associated commands are used to define
%% the authors and their affiliations.
%% Of note is the shared affiliation of the first two authors, and the
%% "authornote" and "authornotemark" commands
%% used to denote shared contribution to the research.
% \author{Anonymous authors}
% \email{trovato@corporation.com}
% \orcid{1234-5678-9012}
\author{Weiwei Jiang}
\email{weiwei.jiang@student.unimelb.edu.au}
\affiliation{%
  \institution{The university of Melbourne}
  \country{Australia}
}

\author{Chaofan Wang}
\email{chaofanw@student.unimelb.edu.au}
\affiliation{%
  \institution{The university of Melbourne}
  \country{Australia}
}

\author{Zhanna Sarsenbayeva}
\email{zhanna.sarsenbayeva@unimelb.edu.au}
\affiliation{%
  \institution{The university of Melbourne}
  \country{Australia}
}

\author{Andrew Irlitti}
\email{andrew.irlitti@unimelb.edu.au}
\affiliation{%
  \institution{The university of Melbourne}
  \country{Australia}
}

\author{Jarrod Knibbe}
\email{jarrod.knibbe@unimelb.edu.au}
\affiliation{%
  \institution{The university of Melbourne}
  \country{Australia}
}

\author{Tilman Dingler}
\email{tilman.dingler@unimelb.edu.au}
\affiliation{%
  \institution{The university of Melbourne}
  \country{Australia}
}

\author{Jorge Goncalves}
\email{jorge.goncalves@unimelb.edu.au}
\affiliation{%
  \institution{The university of Melbourne}
  \country{Australia}
}

\author{Vassilis Kostakos}
\email{vassilis.kostakos@unimelb.edu.au}
\affiliation{%
  \institution{The university of Melbourne}
  \country{Australia}
}

% \author{Julius P. Kumquat}
% \affiliation{%
%   \institution{The Kumquat Consortium}
%   \city{New York}
%   \country{USA}}
% \email{jpkumquat@consortium.net}

%%
%% By default, the full list of authors will be used in the page
%% headers. Often, this list is too long, and will overlap
%% other information printed in the page headers. This command allows
%% the author to define a more concise list
%% of authors' names for this purpose.
\renewcommand{\shortauthors}{Jiang, et al.}

%%
%% The abstract is a short summary of the work to be presented in the
%% article.
\begin{abstract}
  We present a technique to embed information invisible to the eye inside 3D printed objects. The information is integrated in the object model, and then fabricated using off-the-shelf dual-head FDM (Fused Deposition Modeling) 3D printers. Our process does not require human intervention during or after printing with the integrated model. The information can be arbitrary symbols, such as icons, text, binary, or handwriting. To retrieve the information, we evaluate two different infrared-based imaging devices that are readily available -- thermal cameras and near-infrared scanners. Based on our results, we propose design guidelines for a range of use cases to embed and extract hidden information. We demonstrate how our method can be used for different applications, such as interactive thermal displays, hidden board game tokens, tagging functional printed objects, and autographing non-fungible fabrication work.
\end{abstract}

%%
%% The code below is generated by the tool at http://dl.acm.org/ccs.cfm.
%% Please copy and paste the code instead of the example below.
%%
% \begin{CCSXML}
% <ccs2012>
%  <concept>
%   <concept_id>10010520.10010553.10010562</concept_id>
%   <concept_desc>Computer systems organization~Embedded systems</concept_desc>
%   <concept_significance>500</concept_significance>
%  </concept>
%  <concept>
%   <concept_id>10010520.10010575.10010755</concept_id>
%   <concept_desc>Computer systems organization~Redundancy</concept_desc>
%   <concept_significance>300</concept_significance>
%  </concept>
%  <concept>
%   <concept_id>10010520.10010553.10010554</concept_id>
%   <concept_desc>Computer systems organization~Robotics</concept_desc>
%   <concept_significance>100</concept_significance>
%  </concept>
%  <concept>
%   <concept_id>10003033.10003083.10003095</concept_id>
%   <concept_desc>Networks~Network reliability</concept_desc>
%   <concept_significance>100</concept_significance>
%  </concept>
% </ccs2012>
% \end{CCSXML}

% \ccsdesc[500]{Computer systems organization~Embedded systems}
% \ccsdesc[300]{Computer systems organization~Redundancy}
% \ccsdesc{Computer systems organization~Robotics}
% \ccsdesc[100]{Networks~Network reliability}

%%
%% Keywords. The author(s) should pick words that accurately describe
%% the work being presented. Separate the keywords with commas.

%%
%% This command processes the author and affiliation and title
%% information and builds the first part of the formatted document.
\maketitle

% Teaser figure.
\begin{figure}[t]
    \centering
    \includegraphics[width=0.75\textwidth]{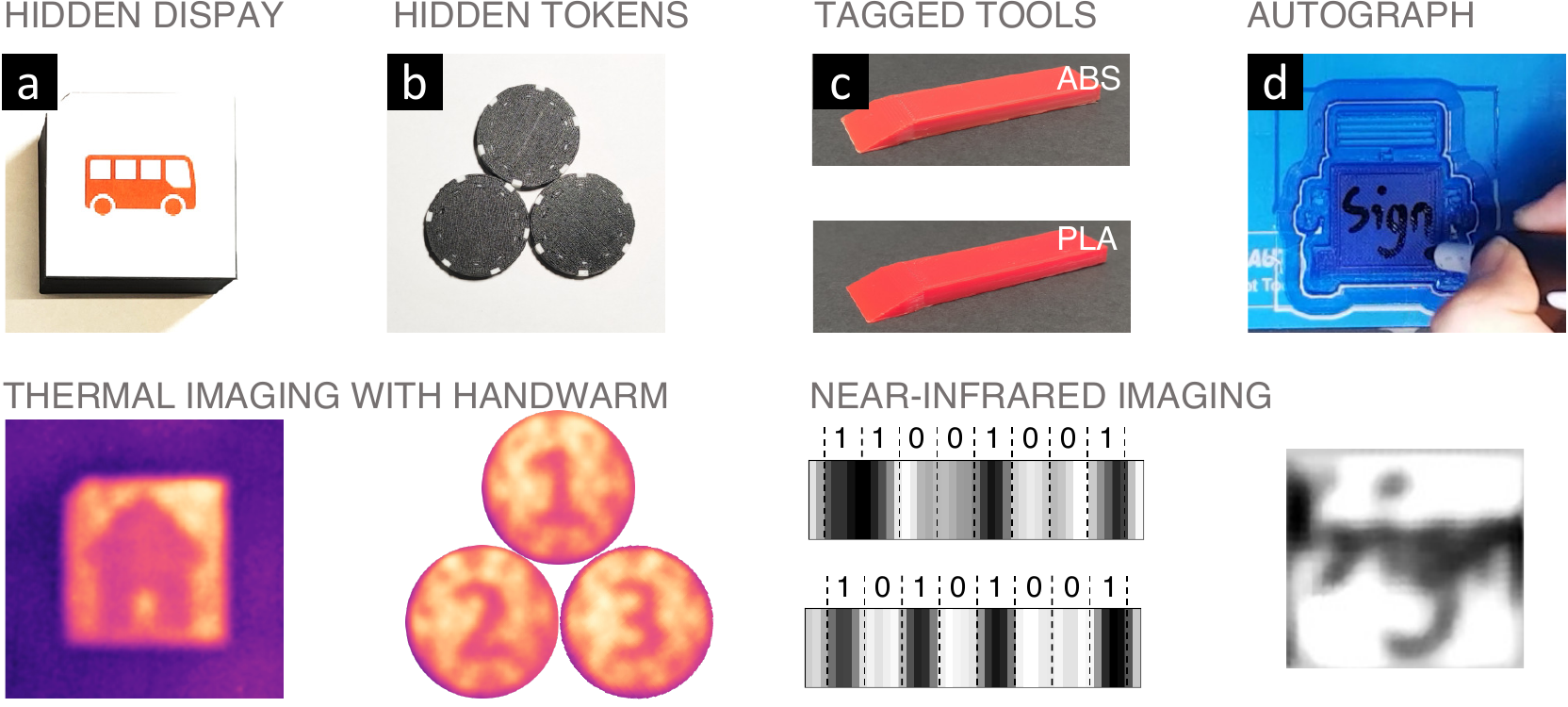}
    \caption{Examples of \textit{\systemname}: \textbf{(a)} A 3D printed hidden display with information both on and hidden under the surface. \textbf{(b)} Three identical models with hidden tags for social activities such as board games. \textbf{(c)} Two 3D printed tools using different materials with metadata (binary tags) under the top surfaces. The tags are durable with heavy use. \textbf{(d)} Autographing inside a 3D printed object. The autograph cannot be erased or modified as it is inside the object.}
    \label{fig:teaser}
\end{figure}

\section{Introduction}
It is increasingly common to come across tagged objects in our daily life. The tags can be text, icons, barcodes or QR-codes printed or attached on an object's surface. Using a tag is an effective way to enable object identification and augmentation. It can also help to link a physical object to a database with further information. This kind of augmentation is particularly helpful in providing design information for digitally fabricated objects, since they are often customized or personalized ~\cite{maia2019layercode, ettehadi2021documented}. 

Recent works show several promising ways to embed tags within digitally fabricated objects. Existing methods, however, have several issues. A common approach is to design a tag on the surface. This method can easily compromise the aesthetics of the object~\cite{maia2019layercode}. In comparison, the object design can be modified by extrusion or engraving for tagging~\cite{ettehadi2021documented}; however, this can also be obtrusive, and can limit the object's functionality. Alternatively, recent work has shown promising results in embedding information under the surface, using high-end or professional 3D printers (\eg PolyJet)~\cite{willis2013infrastructs,li2017aircode}. To date, however, such techniques have been unachievable with consumer-level FDM (Fused Deposition Modeling) printers, which are affordable, accessible and more suitable for many everyday use cases~\cite{shewbridge2014everyday}. To this end, we still lack an effective FDM 3D printer based tagging method that is printable with the object, and does not compromise its design aesthetics.

We present \textit{\systemname}, a technique for invisibly embedding information (\eg tags) into 3D printed objects using FDM 3D printers. The embedded information is unobtrusive or invisible to the eye. Our technique does not alter the surface design of the object, allowing for more creative and secure designs. With \textit{\systemname}, it is possible to simply design and print information under the surface, without human intervention during or after printing.

To read the information, we leverage two methods for different use cases. One method is based on the different thermal conduction rate between the air and the 3D printed material. This is possible as most objects printed with an FDM 3D printer consist of an infill pattern with the majority of the space being filled with air. The information under the surface can be revealed using a thermal camera after applying heat to the surface (\eg by hand-warming, as shown in Figure~\ref{fig:teaser}a and b). The second method is based on the fact that many plastic materials can be penetrated by or reflect near-infrared light~\cite{reich2005near}. By printing the information and the object using different materials or colors, the information can be read using a low-cost near-infrared device (Figure~\ref{fig:teaser}c and d). 

Our contribution is three-fold. First, we present a technique to embed information into a 3D printed object using consumer-level FDM 3D printers. Our method expands the design space for tagging 3D printed objects without compromising their appearance or shape. Second, we provide comprehensive design guidelines for embedding information into 3D printed objects addressing multiple application scenarios and use cases. Third, we provide several example applications using an embedded information scheme, including a novel way to interact with digitally fabricated objects using thermal cameras, and a secure and non-fungible way to autograph a 3D printed object as a unique digital fabrication work. 

\section{Related work}
\subsection{Tagging 3D printed objects}
% On-the-surface tagging methods.
\subsubsection{Tagging on the surface}
\hfill \\
An intuitive and convenient way to tag an object is to attach a label on the surface. For 3D printed objects, such a label can also be printed as a part of the design or fabrication process~\cite{ettehadi2021documented}. For example, Harrison~\etal proposed a notch pattern to tag a 3D printed object's surface. The notches can be swiped to generate a complex sound, which is then recorded by a microphone and decoded to a binary ID~\cite{harrison2012acoustic}. Similar work using a comb-like structure was presented by Savage~\etal~\cite{savage2015lamello}. Besides acoustics, Maia~\etal presented LayerCode, a scheme to embed optical barcodes by varying the colors of different layers, using a dual-color 3D printer~\cite{maia2019layercode}. However, LayerCode alters the appearance of the whole object, which can be undesirable. Although the authors demonstrated a method to make the information invisible, utilizing a customized material mixed with near-infrared dye and a modified 3D printer made the approach unrealistic for daily use. To address this issue, Delmotte~\cite{delmotte2020blind} proposed a less restrictive method by varying the surface layer thickness. The authors developed a tool to manipulate the G-Code for tagging a 3D printed object on the surface. However, the area with the tag resembles printing flaws, and may be removed by daily use or during post-processing (\eg polishing). This limitation also applies to the tagless method that utilizes subtle print artifacts or patterns for 3D printed object recognition~\cite{dogan2020gid}. 

% Textile
In addition to binary tags mentioned above, existing works also show different ways to tag 3D printed objects with other textures. In particular, recent studies focus on incorporating textiles or fabrics in 3D objects to further expand the design space using FDM 3D printers with not only more applications, but also tagging methods by embedding textures on the printed objects. For instance, Rivera~\etal demonstrated a technique to embed textiles into 3D printed objects, which can leverage well-developed embroidery techniques as a potential way for tagging~\cite{rivera2017stretching}. Besides embedding, Takahashi~\etal presented a method to print textiles using a 3D weaving technique~\cite{takahashi2019printed}. The authors demonstrated that the fabrics can be printed using an FDM 3D printer by controlling the printer's head movements~\cite{takahashi2019printed}. Alternatively, Forman~\etal showed that such 3D printed quasi-fabrics could also be printed by under-extrusion~\cite{forman2020defexfiles}. Both works showed that 3D printed textiles with patterned textures can potentially be used as visual tags on the surface. It is worth noting that textile-like methods are mostly suitable for flexible objects, and may not be appropriate for long-term or heavy use. 

% Touch-screen
Beyond visual patterns, researchers in the HCI community also developed methods to tag conductive patterns on the surface that can be read by a touchscreen. For example, Marky~\etal showed a design to embed a conductive pattern on the bottom surface of an object. The pattern can be re-configured by the user, and thus, can be used for two-factor authentication in a tangible way~\cite{marky2020auth}. Similarly, Schmitz~\etal presented Itsy-Bits, a fabrication pipeline to design small footprints for detecting tangibles on a capacitive touchscreen~\cite{schmitz2021itsybits}. However, their methods are limited to specific patterns and require user contact while reading, which may not be adopted to general information embedding applications.

Even though the use of `on-the-surface' tagging methods is beneficial in several ways, such as convenience of labeling or intuition for reading, there are several important drawbacks. In particular, such a method can alter the appearance, shape or functionality of the object, and is less robust for long-term or frequent use. 

% Under-the-surface tagging method. 
\subsubsection{Tagging under the surface}
\hfill \\
In contrast to on-the-surface tagging methods, tagging an object under the surface is less intuitive or convenient to read. This method does not compromise the external design of the object, while also being more enduring. A common approach involves embedding an RFID tag inside a digitally fabricated object~\cite{spielberg2016rapid}. However, such a method requires additional materials and procedures to perform embedding, as a fully printable technique is not currently possible~\cite{espalin20143d}. Our work addresses this issue and enables embedding information (\ie tagging) within a 3D printed object that is printable using an off-the-shelf FDM 3D printer. 

A study by Li~\etal is most closely related to our work~\cite{li2017aircode}. The authors presented AirCode, a technique to embed a QR-Code like pattern under the surface of a 3D printed object. The code can be printed using a PolyJet 3D printer with well-designed cavities inside the 3D model. The code can then be read by a monochrome camera and a projector, and decoded as the tag of the printed object (\eg metadata or ID). However, AirCode cannot be used with the common consumer FDM 3D printer. This is because FDM 3D printers yield non-homogeneous printouts, and use relatively thick materials compared to an expensive PolyJet 3D printer. Also, AirCode only embeds binary data, and requires assembly after printing, since a PolyJet 3D printer cannot print hollows (cavities) inside an object. Another similar work is InfraStruts~\cite{willis2013infrastructs}, an information embedding scheme for layered structures, allowing the embedding of not only binary data, but also icons and text. However, their method is also limited to PolyJet 3D printers and is susceptible to layer variations such as uniformed thickness, as the authors use a THz-TDS (TeraHertz Time-Domain) device for imaging. While there are prior examples of using FDM 3D printers to embed information, their design severely limits functionality in application and durability~\cite{silapasuphakornwong2015nondestructive,suzuki2017information}. Their methods are incapable of producing embedded information beyond binary due to their restrictive design between adjacent bits. Our approach does not have this limitation, and can produce binary, text and icons using a more accessible setting. Other works using highly customized materials make the settings unrealistic for daily use~\cite{silapasuphakornwong2018information,silapasuphakornwong20193d,silapasuphakornwong2019technique}.

% Electronics
Beyond embedding information for imaging, recent works also include fabricating information as signals inside an object. For instance, Iyer~\etal demonstrated a backscattering system that enables wireless analytics for 3D printed objects~\cite{iyer2018wireless}. The authors embedded antennas that can be switched on or off using a sophisticated mechanical design, triggered by user interactions. The information can then be captured by reading the signals backscattered by the antennas. In another example, Chadalavada~\etal presented ID'em~\cite{chadalavada2018id} that allowed users to embed a conductive dot matrix under the surface that can be read by an array of inductive sensors. The conductive matrix works like an antenna array that manipulates the magnetic fields as the signals. The signals are then decoded as the information. However, both of the aforementioned methods are limited to specific data types (categorical or binary), and may not be easily printed using common off-the-shelf 3D printers. 

\subsection{Embedding objects with 3D prints}
In a broad sense of information embedding, significant efforts have been done for embedding physical objects within 3D prints. In particular, researchers have focused on incorporating electronics with 3D printed objects that enable functionality or interactivity, \eg camera~\cite{savage2013sauron}, a wireless accelerometer~\cite{hook2014making}, or a speaker~\cite{ishiguro2014printed}. A recent study also demonstrated novel use cases enabling 3D printing on cloth with SMA (Shape-Memory Alloys) actuators~\cite{muthukumarana2021clothtiles}. Furthermore, researchers endeavor to provide tools and guidelines for designing inner structures to embed electronics, \eg for designing internal pipes~\cite{willis2012printed, savage2014series} or hollow tubes~\cite{tejada2020airtouch}, optimizing sensor placement~\cite{bacher2016defsense}, component placement for assembly~\cite{desai2018assembly}, or even for printed objects that are deformable~\cite{wang2020morphingcircuit, hong2021thermoformed, ko2021designing} or re-configurable for fast-prototyping~\cite{schmitz2021ohsnap}. However, such methods are not fully printable yet, and cannot be automated. Therefore, it still requires relatively complex fabrication techniques with extra material costs, which is extravagant for simple tagging or information embedding tasks.

% Print conductive traces or surfaces 
One way to overcome this non-printable disadvantage is to use the emerging conductive 3D printing materials. Recent works show potential to print conductive traces, circuits or surfaces within objects. For example, Schmitz~\etal presented a design pipeline to print selected surfaces using conductive ABS (Acrylonitrile Butadiene Styrene) for sensing touched areas on a 3D printed object~\cite{schmitz2015capricate}. As an extension, the authors further showed a pipeline for embedding electrodes to fabricate a hover-, touch-, and force sensitive object~\cite{schmitz2019trilaterate}. Besides embedding electronics with the traces, such a method can also be useful for various interactive scenarios in VR and AR~\cite{narazani2019extending}. Moreover, such a method can be adopted for heavily used functional objects utilizing emerging carbon fiber strengthened 3D print materials. For example, Swaminathan~\etal demonstrated a technique to fabricate conductive traces while keeping the object mechanically strong for heavy use~\cite{swaminathan2019fiberwire}. In addition to solid conductive traces, recent work has adopted conductive liquids for fabricating circuits using a 3D printed stamp as the circuit mold~\cite{tokuda2021flowcuits}. Beyond rigid objects, existing works also show the feasibility of embedding conductive traces or electronics for 3D printed flexible objects, such as conductive fabrics~\cite{peng2015layered}, electrospun textiles~\cite{rivera2019desktop}, using foldable structures~\cite{olberding2015foldio, yamaoka2019foldtronics}, or the flexible TPU (Thermoplastic Polyurethane) material with silver inks (\ie AgTPU)~\cite{valentine2017hybrid}. In particular for the flexible objects, researchers have strived to design a computational structure for enabling functionality or information embedding. For example, Schmitz~\etal designed a flexible structure fabricated using TPU that can sense deformation as input information~\cite{schmitz2017flexibles}. Deformation structures can also be used with pneumatic devices for sensing user inputs~\cite{vazquez2015printing,he2017squeezapulse}. Nevertheless, existing methods are not fully printable and still require electronics to be attached~\cite{macdonald2016multiprocess}. 

% Liquids 
Beyond electronics, previous works also show other promising methods to embed everyday objects for 3D prints. In particular, Schmitz~\etal demonstrated techniques to embed liquids inside 3D printed object as a changeable information, for sensing tasks such as tilting and motion~\cite{schmitz2016liquido, schmitz2018offline}. Such a method can also be beneficial for reducing cost of fabrication time or material, and has a great potential for information embedding~\cite{mueller2014fabrickation, chen2018medley, wall2021scrappy, sun2021shrincage}. Other than 3D printing, researchers also proposed different digital fabrication based methods for information or object embedding, such as 3D sculpting~\cite{oh2018pep}, laser-cutting~\cite{takahashi2019printed, valkeneers2019stackmold}, paper printing~\cite{li2016paperid}, or spraying~\cite{hanton2020protospray}.

In summary, existing works for embedding information inside a 3D printed object are not practical for general information embedding; they either require non-printable materials or cannot be applied to off-the-shelf consumer FDM 3D printers. 

% Briefing how our system works.  
\section{Method overview}
\label{sec:method}
We present a technique to embed information under the surface of 3D printed objects. The embedded information itself is printable as part of the printed objects, and cannot be directly seen to the eye. In this section, we briefly describe the method to design, read, and fabricate. 

\subsection{Information embedding}
\begin{figure}[t]
    \centering
    \includegraphics[width=0.9\textwidth]{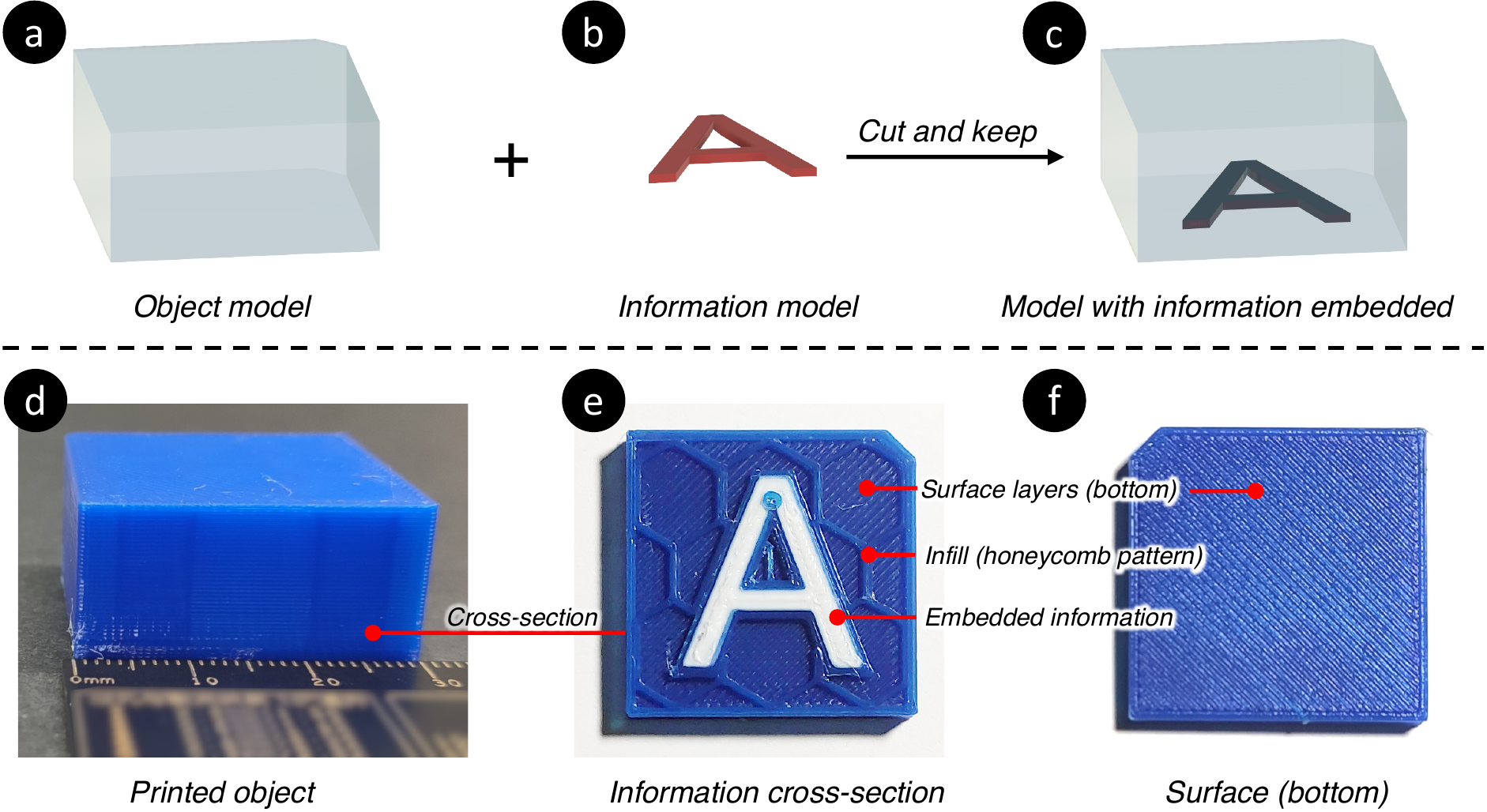}
    \caption{Illustration of the design overview. \textbf{(a)} An object model to embed information inside. \textbf{(b)} The information model to be embedded. \textbf{(c)} The object model with information embedded. \textbf{(d)} The 3D printed object with information embedded. \textbf{(e)} The cross-section with information embedded. \textbf{(f)} The surface under which the information is embedded.}
    \label{fig:design_overview}
\end{figure}

Our design pipeline is shown in Figure~\ref{fig:design_overview}. In this paper, we use Autodesk Fusion 360~\cite{fusion360} as the 3D modeling tool, without requiring any modification or plugin. Thus, the pipeline can also be applied to other 3D modeling software. The whole design pipeline includes three steps:

\begin{enumerate}
    \item \textit{Object modeling}: Model the desired object. For a mesh file (\eg stl file), Fusion 360 provides a ``Mesh to BRep'' function that allows readily editing the mesh file. 
  
    \item \textit{Information modeling}: Model the necessary information (\eg letter 'A'). The information should be sketched and extruded as a 3D object. 
    
    \item \textit{Information Embedding}: Embed the information model inside the object model. Here, we embed the model for ``A'' into the object. Then the object model is subtracted by the ``A'' model, while keeping the ``A'' model in place. In Fusion 360, this can be achieved by invoking the ``combine'' command. For a mesh editor software, this can be achieved with ``subtractive boolean'' (\eg Blender~\cite{blender}). 
\end{enumerate}

The designed model must be exported individually. In Fusion 360, the export option is ``one file per body'' for fabrication. This allows the slicing software to have different settings (such as materials) for the object model and the information model. For example, we demonstrate the fabricated object in Figure~\ref{fig:design_overview} (d)-(f), where the object is printed using PLA-Blue, while the information model is printed using PLA-White. 

% 3D modeling.

\subsection{Reading principles}
\label{subsec:principles}
\begin{figure}
    \centering
    \includegraphics[width=0.9\textwidth]{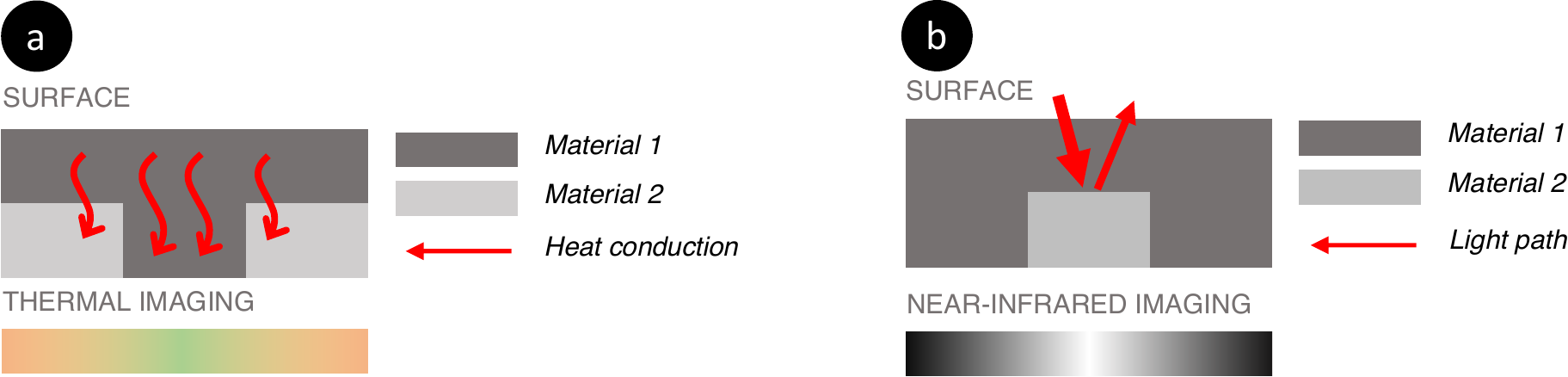}
    \caption{Illustrations of the two reading approaches. Bottom part demonstrates the imaging patterns.}
    \label{fig:imaging_principles}
\end{figure}

Since we aim to embed information that cannot be directly seen on the surface of the object, the information needs to be imaged (visualized) for reading. In this paper, we leverage the following two approaches for imaging: 

\begin{enumerate}
    \item \textit{Thermal conduction}: As illustrated in Figure~\ref{fig:imaging_principles}~(a), after heat is transferred to the surface of an object, the heat flows from that surface down into the object. Because the object's inside is composed of two different materials, the heat permeates at different speeds, depending on thermal conduction properties of the materials. As a result, the temperature of the surface changes at different speeds, resulting in a thermal pattern as a projection of the information model on the surface. Here, we consider the filament materials (\eg PLA or ABS) as one such material, while air is the other material (the spaces without infill). 
    
    \item \textit{Light penetration and reflection}: As illustrated in Figure~\ref{fig:imaging_principles}~(b), a 3D printed object can also be fabricated by two materials with different optical characteristics. In particular, we are interested in the case where the material on the surface layers can be penetrated while the other material inside the object is reflective. By transmitting light onto the surface of the object and measuring the reflected light, we can create an image of the reflection pattern. Further, as we aim to embed information unobtrusively, we utilize near-infrared lights that are more capable of penetrating and distinguishing different materials, compared to visible lights~\cite{reich2005near}. For 3D printing, we can use filaments of two different colors (such as PLA-blue and PLA-white). 
\end{enumerate}

In this paper, we demonstrate how to read the embedded information using a thermal camera and a near-infrared scanner respectively. We use the Optris Xi 400\footnote{\url{https://www.optris.com/optris-xi-400}} thermal camera, with wavelengths $8~\mu m$ -- $14~\mu m$ (mid-far infrared). We also use the DLP NIRscan Nano~\cite{dlpnirscanonline} near-infrared scanner. For raster scanning, we mount the near-infrared scanner on an xy-plotter, controlled by a custom-built software. We also provide a supplementary video demonstrating our system with both reading devices.

\subsection{Fabrication methods}
We propose one fabrication technique for thermal imaging, and another for near-infrared imaging (Figure~\ref{fig:imaging_principles}). Both fabrication methods use the same design models (\ie the same stl files for slicing). 

% 3D printing.
\begin{figure}[t]
    \centering
    \includegraphics[width=\textwidth]{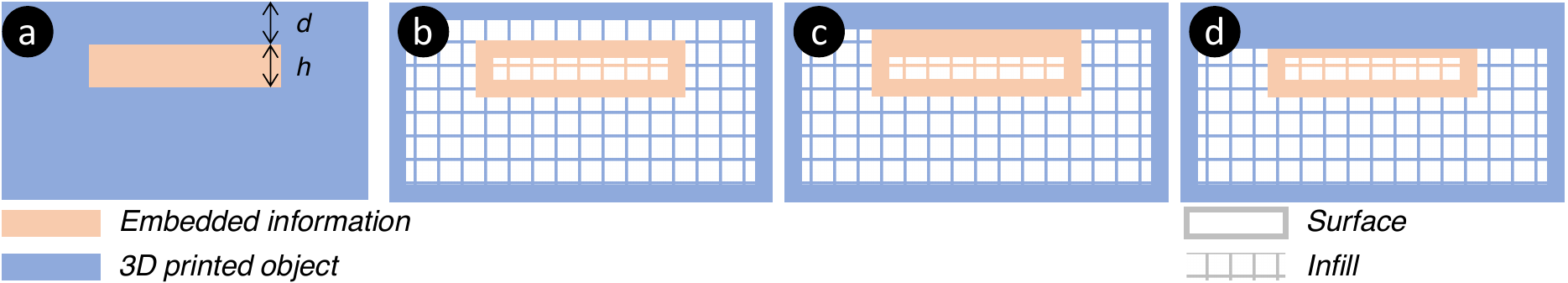}
    \caption{Fabrication methods. \textbf{(a)} Illustration of the vertical cross-section. \textbf{(b)} Normal fabrication method. \textbf{(c)} The ``surface-join'' fabrication method. \textbf{(d)} The ``surface-fill'' fabrication method.} 
    \label{fig:fabrication_overview}
\end{figure}

\begin{enumerate}
    \item \textit{Surface-join, for thermal imaging}: As illustrated in Figure~\ref{fig:fabrication_overview}~(c), the ``surface-join'' denotes the surface layers of the object and the embedded information model are joined together. In slicing software, this is done by thickening the top or bottom layer of both the object model and the information model. This enables the thermal conduction between the object's surface layers and the information model. 
    
    \item \textit{Surface-fill, for near-infrared imaging}: As illustrated in Figure~\ref{fig:fabrication_overview}~(d), the ``surface-fill'' denotes the surface layers of the object are filled until the top of the 3D model representing the information. In slicing software, this is done by thickening the surface layer of the object model only. This is for near-infrared imaging and to prevent light scattering due to the non-uniformed infill material (filament and air) between the surface layers and the top of information model for reflection. 
\end{enumerate}

We note that our two methods are exclusive. Depending on specific applications and use cases, either method should be chosen for fabrication. We show various examples in the following section and provide comprehensive design guidelines in Table~\ref{tab:design_space}.

% Show the examples, this will take more than 1/4 contents of the paper.  
\section{Example Applications}
\label{sec:examples}
\textit{\systemname} enables various applications by embedding information inside 3D printed objects using off-the-shelf FDM 3D printers and filaments, without requiring extra procedures (such as assembly), or software modifications. We exemplify the applications including tagging 3D models, interactive thermal displays, and hand-signing (autographing) 3D prints. We present the design details in Section~\ref{subsec:design_guideline} below. We also demonstrate our application in the supplementary video. 

\begin{figure}[t]
    \centering
    \includegraphics[width=\textwidth]{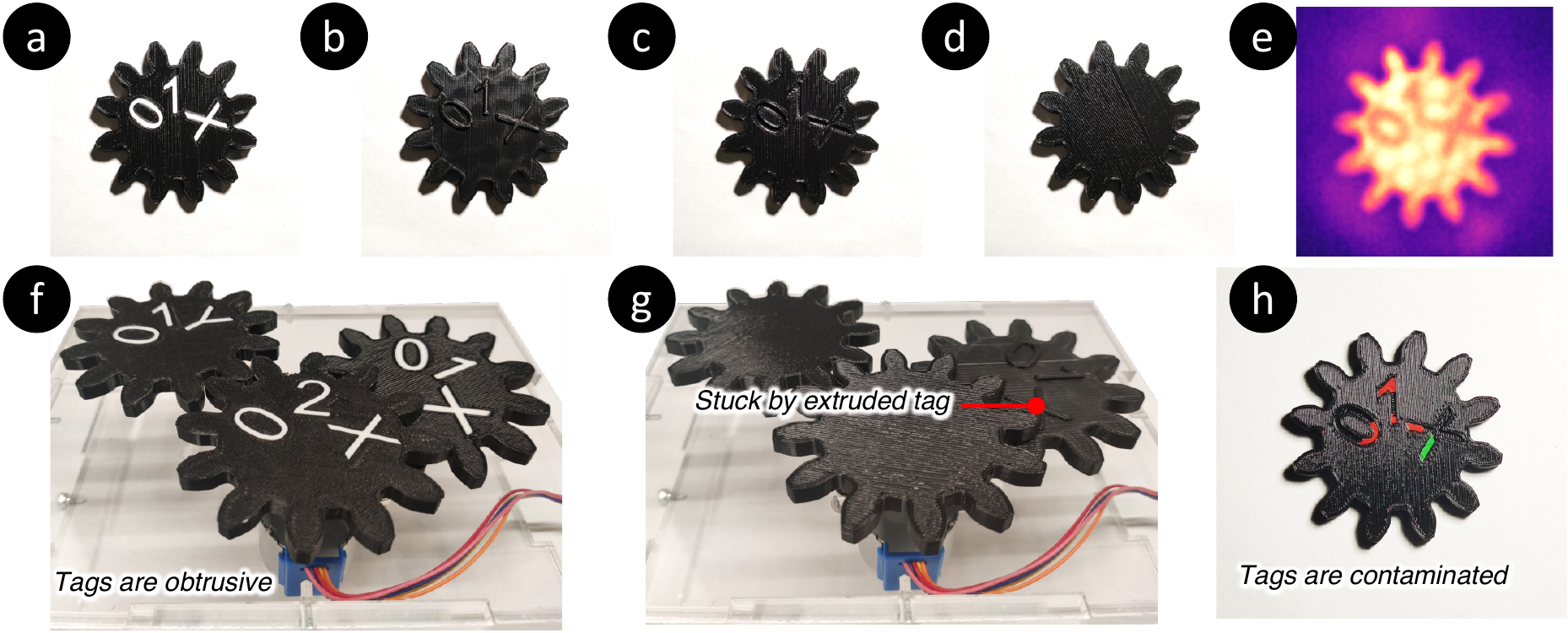}
    \caption{Demonstration of different tagging methods by printings. The example is to 3D print a tag for a gear with the part number (01X). \textbf{(a)} The tag is printed on the surface using another color. \textbf{(b)} The tag is printed above the surface with extrusion upwards. \textbf{(c)} The tag is printed by extruding the tag inwards (like engraving). \textbf{(d)} The tag is printed under the surface. \textbf{(e)} The tag is read under the thermal camera for the object in (d), after the surface is warmed by hand. \textbf{(f)} Illustration of multiple gears with labels on the surface. \textbf{(g)} Illustration of malfunctioning gears tagged by extrusion in (b). \textbf{(h)} Illustration of visual contaminated tagged in (c).}
    \label{fig:demo_gears}
\end{figure}

\subsection{Tagging 3D models}
Conventional tagging methods for 3D printed objects are usually performed on the surface of the object. Such a method can alter not only the appearance of the object, but also constrains its shape or even functionality (Figure~\ref{fig:demo_gears}). In particular, for comparison, we include examples using the most common conventional tagging methods: (a) Tagging on the surface using another color. (b) Tagging by extruding outwards. (c) Tagging by extruding inwards (like engraving). Furthermore, we consider two types of tags as seen below.

\subsubsection*{Serial number: } Serial numbers can be used to tag different objects with similar designs. For the examples shown in Figure~\ref{fig:demo_gears}f and g, we include three gears composed in a gearbox, tagged as ``01X'', ``02X'' and ``01Y''. The gearbox itself can be used as part of mechanical design, educational activity or exhibition. For example, in an education setting, \textit{\systemname} could let students try and figure out the correct placement of the gears, whilst also embedding hidden information as hints for the students to read in case if they need help. Compared to other methods, our method is unobtrusive and does not affect the design functionality.

\begin{figure}[t]
    \centering
    \includegraphics[width=\textwidth]{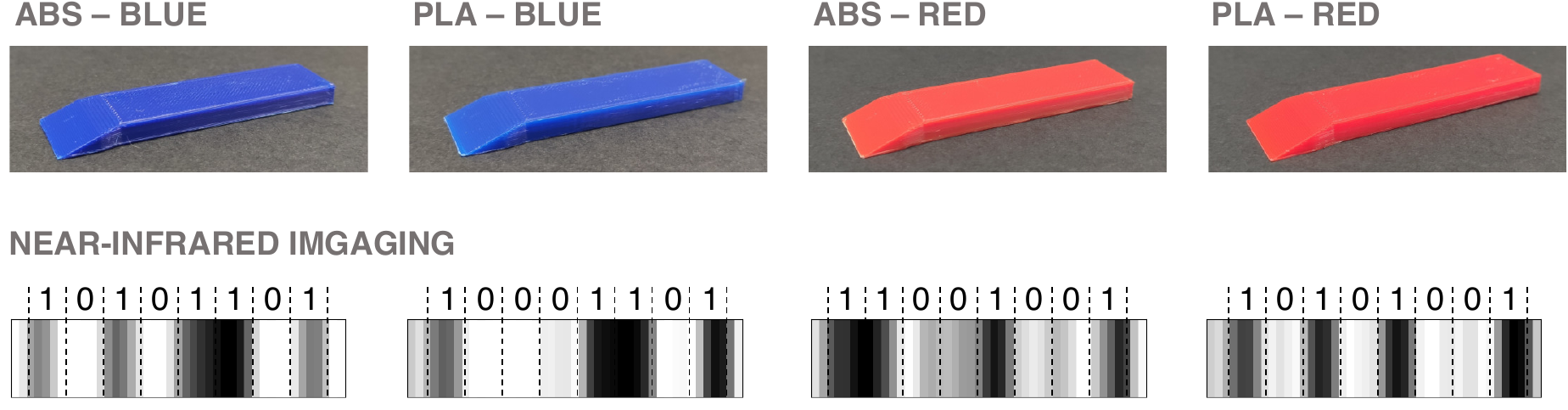}
    \caption{Examples of tagging 3D printed tools with binary data. Embedded tags are robust for frequent and heavy use, compared to tags on the surface. }
    \label{fig:demo_tools}
\end{figure}

\subsubsection*{Binary data: } Binary data can be used to store metadata for the object, \eg the key or ID for a database entry that stores detailed information of a print or the design. For example, as shown in Figure~\ref{fig:demo_tools}, a tool can be printed using different materials (\eg ABS or PLA) or different colors. The objects may not be easily distinguished after being printed. Since a tool can be frequently and heavily used, tags on the surface can be easily damaged (\eg due to scratching or abrasion). In contrast, \textit{\systemname} embeds information under the surface and reduces the likelihood of damage to the tags.  

\begin{figure}[t]
    \centering
    \includegraphics[width=\textwidth]{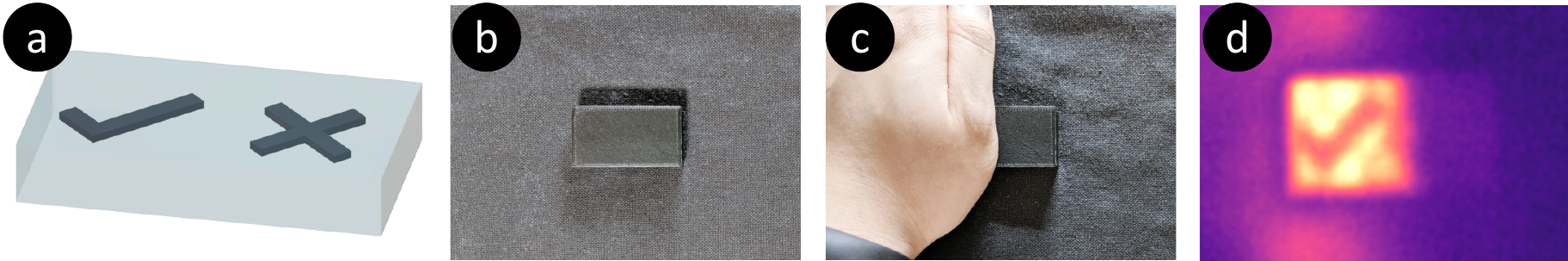}
    \caption{Example of a 3D printed interactive thermal display. \textbf{(a)} The model design with different information embedded on the left part and the right part respectively. \textbf{(b)} The printed object with information embedded. \textbf{(c)} Selectively hand-warm the left part of the object. \textbf{(d)} The information on the left side is imaged under the thermal camera, while the information on the right side remains hidden.}
    \label{fig:demo_display}
\end{figure}

% Continue the previous interaction example. 
\subsection{Interactive thermal displays}
To read the embedded information, one of the approaches we leverage is thermal conduction (Section~\ref{subsec:principles}). The information can be imaged after applying heat on the surface, \eg through warm hands. This enables a new way of interacting with a 3D printed object. As demonstrated in Figure~\ref{fig:demo_display}, we embed two pieces of information into an object. The information can be selectively revealed by interacting with a specific part of the object (left or right), while the other part remains hidden. 

\subsubsection*{Security code. } In practice, such an interaction can be used for security purposes. For example, most credit cards have a CVV2 (Card Verification Value 2) code on the back side. Such a number is used for reducing credit card fraud when the card is not present for transaction~\cite{gupta2017credit}. However, the CVV2 code is usually printed, affording easy malicious access for potential data breaches. In similar design, the grid authentication, where the user has a grid card containing a matrix of codes for two-factor authentication is also subject to data breach risks, since the grid card codes are visible and can be photographed~\cite{ali2019consumer}. \textit{\systemname} could address the above mentioned issues as it hides the secret information (\ie codes) under the surface. The codes can be revealed for a short period of time (detailed in Section~\ref{subsec:eval_thermal}) by interacting with the object whenever required. In addition, the location of information can be known only to the user, further improving the security of the data.

% Continue the previous example. 
\subsubsection*{Hidden tokens for social activities. }
\textit{\systemname} can also can be used in social settings such as escape rooms and board games, with temporal uncovered information. For example, objects with hidden tokens can be placed or installed in an escape room as concealed clues of puzzles. The token can only be read after interacting with the object while holding a thermal camera (\eg recent smartphones are equipped with thermal camera such as~\cite{catphones}, or mobile thermal cameras such as~\cite{flir}). Such a design can also be used for board games, where players can be assigned with different roles using tokens or cards. With \textit{\systemname}, we can create hidden tokens that can only be seen under a thermal camera after interacting with the token. For instance, a player can act as the ``oracle'' holds a thermal camera to identify other players' roles. Potentially, this may provide more possibilities to design such social activities, compared to using visible or covered tokens.

% Continue the previous "secret" and "token" example. 
\subsection{Extension: autographing unique or non-fungible 3D prints}
\begin{figure}[t]
    \centering
    \includegraphics[width=\textwidth]{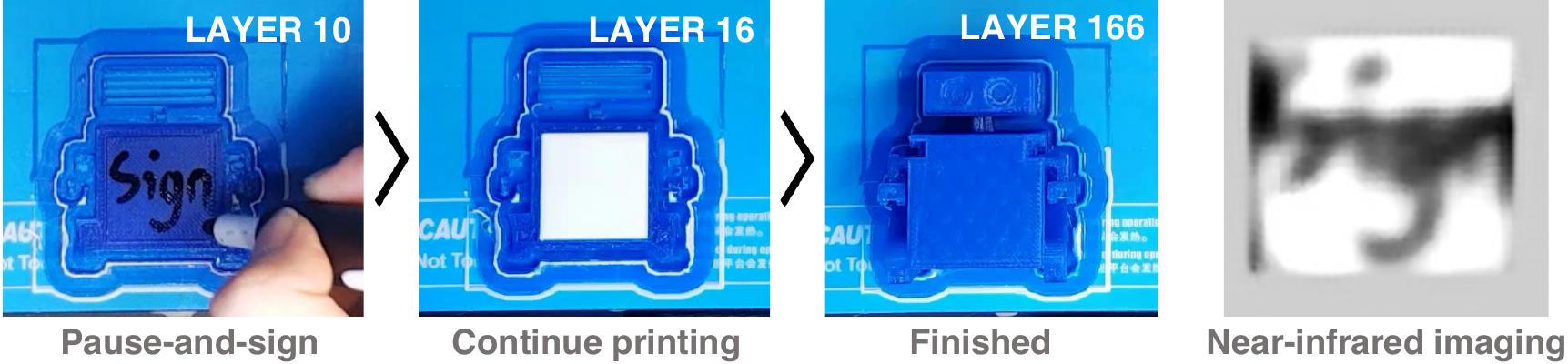}
    \caption{Demonstration of autographing inside a 3D printed object. \textbf{Layer 10}: The autograph is signed by pausing the print job. The job is resumed after autographing. \textbf{Layer 16}: A reflective plate is printed using PLA-White. \textbf{Layer 166}: The job is finished. After printing, the autograph can be imaged by the near-infrared scanner.}
    \label{fig:demo_sign}
\end{figure}

Finally, as an extension to information embedding, \textit{\systemname} enables autographing inside the 3D prints. The 3D prints can then be treated as a unique object (such as an artwork), or a non-fungible object, as the autograph cannot be erased or modified (compared to autograph on the surface). As a demonstration, we show a robot model with an autograph under its back surface in Figure~\ref{fig:demo_sign}. The autograph is then imaged using the near-infrared scanner. Not only can autographs be encoded using \textit{\systemname}, but other hand-written information, \eg text, icons or binary data with precise hand-draw. In addition, the handwritten information (\eg autograph) can be modeled and shared with the object design as a unique digital signature of the design.

% Evaluations, figures, etc. 
\section{Evaluation and Validation}
\label{sec:eval}
Finally, we evaluate our information embedding method to better understand its limitations, including using two different imaging approaches: thermal imaging and near-infrared imaging. A comparison between the two approaches is shown in Table~\ref{tab:thermal_vs_nirs}. Furthermore, based on our results, we derive design guidelines for information embedding and reading using an FDM 3D printer for different use cases. The guidelines are included at the end of this section (Table~\ref{tab:design_space}).

\subsection{Samples} 

For systematic evaluation we adopt information encoding conventions previously used in literature~\cite{willis2013infrastructs, li2017aircode}, and consider a $4\times4$ binary matrix as the embedded information. Such a matrix can also be considered as a bitmap for non-binary data. The binary matrix is randomly generated, with 8 bits being $1$s and the other 8 being $0$s, with the top-left bit fixed as $1$ as the anchor bit (Figure~\ref{fig:sample_groundtruth}). The matrix is embedded into a cube with dimensions $W \times D \times H = 30 \times 30 \times15~mm$. We then fabricate the cube with different print settings, including:

\begin{enumerate}
    \item Information depth ($d$): The minimal distance between the surface and the information surface. As illustrated in Figure~\ref{fig:fabrication_overview}. 
    
    \item Information density ($X$): The block size (\ie the size of each bit or pixel) in the matrix, measured by $mm~per~pixel$.
    
    \item Infill percentage: The infill percentage of the model. Both printer's heads use the same value. We vary the infill percentage as $10\%$, $20\%$, $40\%$ and $80\%$, considering the use case of modeling, standard printing tools, functional tools and heavily used tools, respectively. 
\end{enumerate}

% Compare between NIRS and thermal.
\input{tables/thermal_vs_nirs}

For fabrication, all samples are printed using a low-cost off-the-shelf dual-head FDM 3D printer (FlashForge Creator Pro 2\footnote{\url{https://www.flashforge.com/product-detail/51}}). The information models (\ie the matrix) are printed using PLA-white filaments. The cube models are printed using PLA-black filaments and PLA-blue filaments, for thermal imaging and near-infrared imaging, respectively. The colors are commonly available and chosen based on our preliminary tests of performance. We also validate our results using other colors. We refer to the example applications demonstrated in Section~\ref{sec:examples} above. Other parameters are fixed for feasibility and consistency (for example, we chose honeycomb for an infill pattern, and set the information height to $1~mm$. Nevertheless, we consider these parameters as negligible).

% Reading results for thermal camera. 
\subsection{Thermal imaging}
\label{subsec:eval_thermal}
\subsubsection{Reading process}
\begin{figure}[t]
    \centering
    \includegraphics[width=\textwidth]{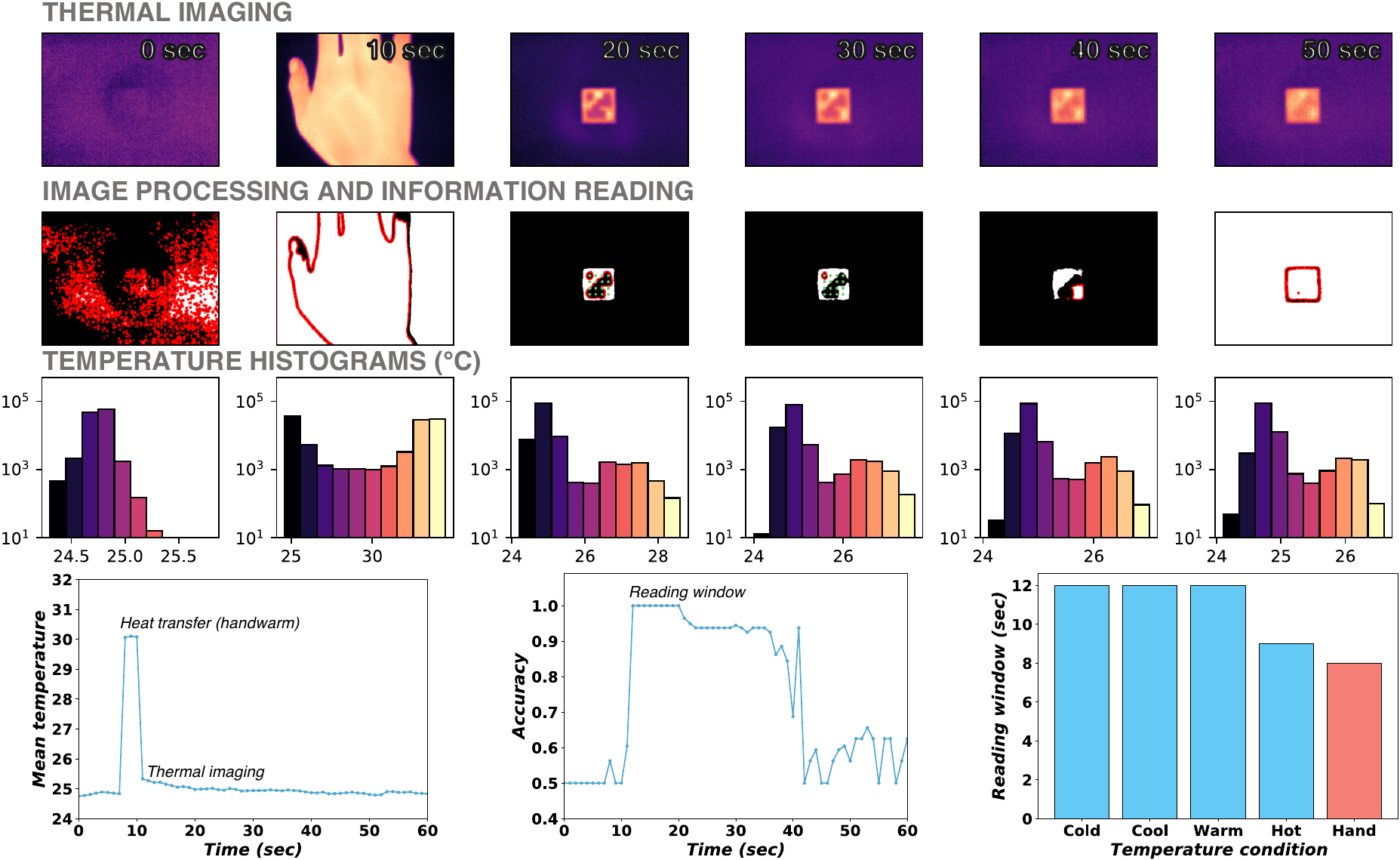}
    \caption{Illustration thermal imaging results. The example demonstrates the hand-warming condition for thermal transfer. The colors in the histograms of different bins represent the colormap for the thermal images. }
    \label{fig:eval_thermal_imaging}
\end{figure}

\hfill \\
We first perform a heat transfer onto the surface of an object embedded with information. The information is revealed when the heat dissipates into the under-surface layers of the object. We record the whole process using the thermal camera and save the data as csv files, at the fastest rate specified by the thermal camera software (5 - 7 frames per second). For each frame, the following two-stage decoding pipeline is performed to decode the matrix into binary data:

\begin{enumerate}
    \item \textit{Normalization}: After reading the frame data, we first denoise the frame using a Gaussian blur filter (window size = $5\times5$), and normalize the values to the range between 0 and 255. 
    
    \item \textit{Binarization 1}: The frame is binarized using Otsu's method. Since the thermal imaging varies in time, we use Otsu's method instead of a selected threshold for better robustness. 
    
    \item \textit{Contour detection 1}: The contour detection algorithm is applied to the binary frame. This step is to detect the contour of the object (\ie cube). 
    
    \item \textit{Cropping}: The normalized frame is cropped using the detected contour with the largest area. 
    
    \item \textit{Binarization 2}: The cropped frame is binarized using Otsu's method. Since Otsu's method is based on the histogram of the image, this second-stage binarization can help the algorithm to better find the threshold and further increase the robustness of the decoding process.
    
    \item \textit{Contour detection 2}: The contour detection algorithm is applied to the cropped binary frame. 
    
    \item \textit{Sampling and decoding}: The center of the top-left contour is used as the anchor point for sampling. A $4 \times 4$ sampling matrix is then derived and decoded to binary data. For evaluation, the spacing between sample points is predefined. 
    
\end{enumerate}
A demonstration of thermal imaging and decoding results are shown in Figure~\ref{fig:eval_thermal_imaging}~(top). We note that the steps above are only for evaluation purposes. In practice, this may not be required for human-readable non-binary information as it does not require decoding (\eg board game token and interactive display). 

\subsubsection{Data collection}
\hfill \\
We first test the imaging results at different temperature conditions. To avoid interference from other parameters, we maintain constant information depth, information density (size) and infill percentage as $1~mm$, $5~mm$ and $10\%$ respectively. Furthermore, considering a practical scenario, we keep the 3D printed object at room temperature ($27\pm2\degree$C). We then prepare four palm-sized bags of sand (made of 100\% cotton calico) at four temperature conditions: ``cold'' ($10\pm2\degree$C), ``cool'' ($20\pm2\degree$C), ``warm'' ($40\pm2\degree$C) and ``hot'' ($50\pm2\degree$C). The sand bags were heated overnight for 12 hours in a temperature-controlled warm-cool dual-mode car fridge to ensure the desired temperature was reached. The temperature errors are caused by the highly dynamic nature of heat dissipation and measurement errors of the thermal camera. Each sand bag is then placed and pressed on top of the cube for three seconds to ensure even and sufficient contact (the surface under which the matrix is embedded). The whole process is recorded by the thermal camera for $60$ seconds, resulting in around $360$ frames collected per condition (varies due to the thermal imaging software). 

In addition to the four above-mentioned controlled conditions, we include an example for practical use -- using a hand to warm-up the cube to reveal the information. Because hand temperature varies for different people at different times of the day~\cite{zontak1998dynamic}, we first regulate the hand temperature using an air-activated hand-warmer. The regulated hand temperature is $35\pm2\degree$C, which is within the normal human temperature range~\cite{hutchison2008hypothermia}. In practice, instead of using an air-activated hand-warmer, this step can be done in other ways, \eg through hand rubbing. 

\subsubsection{Reading window}
\hfill \\
For each frame, the decoding algorithm is executed to detect the binary matrix and calculate the accuracy as the number of bits correctly retrieved divided by 16. Since the thermal camera is very sensitive and requires continuous calibration while recording (performed automatically), a few frames are corrupted before or during calibration~\cite{prakash2006robust}. We exclude those corrupted frames by removing the outliers with performance below the $0.2$ quantile. We note this phenomenon does not impact practical use of the camera as it only affects the frames imminent to the calibration. 

Finally, we calculate the reading window lengths as the duration of error-free decoding (when accuracy = 1.0). The results are shown in Figure~\ref{fig:eval_thermal_imaging}. As an example, we demonstrate the reading process using the hand-warming condition that is more practical than using a sandbag. Evaluation results for all conditions are included in Appendix Figure~\ref{fig:eval_thermal_acc}. We observe that under all conditions, the embedded information can be successfully read immediately after heat transfer. The maximal error-free reading window length is 12 seconds after the sandbag leaves the surface of the cube, achieved by ``cold'', ``cool'' and ``warm'' conditions. The reading window lengths for ``hot'' and ``hand'' conditions are 9 seconds and 8 seconds respectively. This may be caused by the variation of heat dissipation under different conditions~\cite{bergman2011fundamentals}. In particular, the heat dissipates faster with larger temperature differences \footnote{Rate of heat flow = $-kA \cdot \Delta T / \Delta x$, where $k$ is the thermal conductivity, $A$ is the heat emitting area, $\Delta T$ is the temperature difference and $\Delta x$ is the material thickness. In our experiments, $\Delta T$ varies.}. Also, for the ``hot'' condition, the heat dissipates to both the ambient air and into the 3D object.

\begin{figure}[t]
    \centering
    \includegraphics[width=0.8\textwidth]{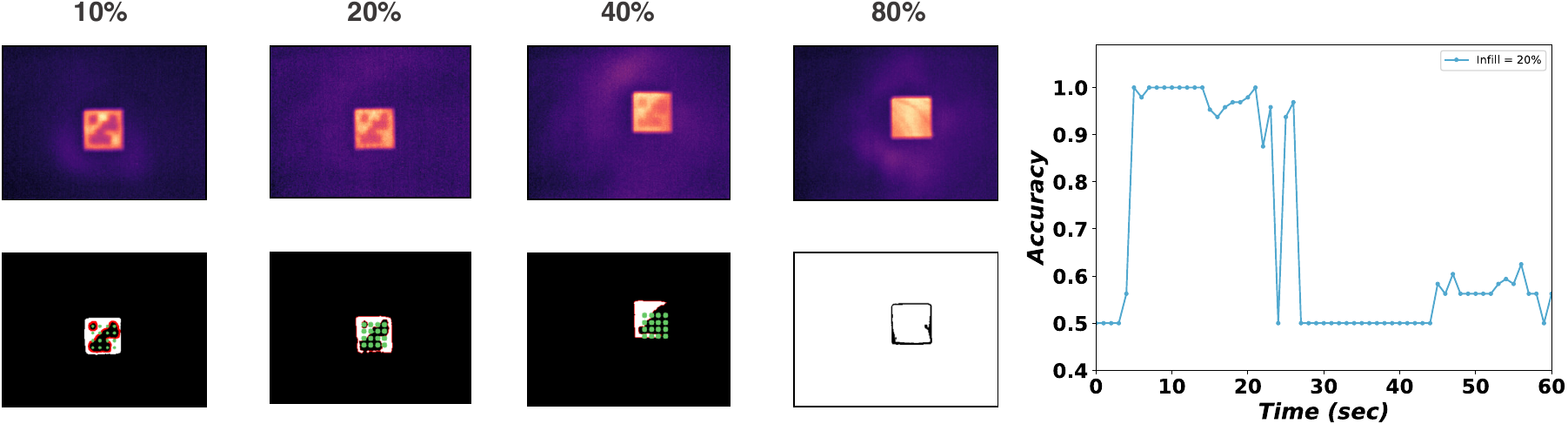}
    \caption{Illustration of thermal imaging with different infill percentages. The figure on the right shows the accuracy in time for $20\%$ infill percentage.}
    \label{fig:eval_thermal_infill}
\end{figure}

\subsubsection{Other parameters}
\hfill \\
\textit{Infill percentage.}
We further test the thermal imaging at different infill percentages. The initial thermal transfer is applied by hand-warming. Our reading succeeds at infill percentage of $20\%$ and fails at $40\%$. The result is expected because a higher infill percentage acquires more heat. The more heat dissipates to the infill prints, the smaller the temperature difference is between the areas embedded with information model and those without, yielding more blurred thermal images. The results are shown in Figure~\ref{fig:eval_thermal_infill}.

\vspace{1em}
\noindent\textit{Information density and depth.}
Our reading fails at higher information density or greater depth (information density $X = 4~mm~per~pixel$ and depth $d = 2~mm$). Similarly, this is limited by the principle of heat dissipation. For a higher information density, the infill prints take relatively more heat. While for the information in a greater depth, the heat dissipates through a longer path to the information model, resulting in more heat loss and less temperature differences between areas embedded with information and those without. The results are illustrated in Appendix Figure~\ref{fig:eval_thermal_w4} and \ref{fig:eval_thermal_d2}.

\subsection{Near-infrared imaging}
\label{subsec:eval_nirs}
\subsubsection{Reading process}
\hfill \\
As briefed in Section~\ref{sec:method}, the samples are raster scanned by a near-infrared scanner mounted on an xy-plotter. For scanning, we set the raster step size as $1~mm$, and home the scanner to the pre-defined area. The scanning resolution is $24\times24$, resulting $576$ near-infrared spectra for each image. The scanner's settings are identical to literature~\cite{klakegg2018assisted,jiang2019probing,jiang2021user} ($900~nm$ - $1700~nm$ wavelength range, $228$ wavelength resolution, $7.03~nm$ light pattern width, and $0.635~ms$ exposure time). As a result, the raw dimension for a near-infrared image is $24\times24\times228$. For each image, the following processing pipeline is performed: 

\begin{enumerate}
    \item \textit{Normalization}: The mean value of each spectrum is computed across the near-infrared wavelengths as the pixel value. Then the image is normalized to the range between 0 and 255. 
    
    \item \textit{Super-resolution}: We adopt a pre-trained deep-learning based super-resolution model to upsample the image by four times~\cite{dong2016accelerating}. Since the raster-scanned resolution is low, this step can enhance the imaging quality, yielding a higher resolution of $96\times96$.
    
    \item \textit{Binarization}: The image is binarized using a simple thresholding method. We empirically select $0.4$ of the maximal value as the threshold for all images. 
    
    \item \textit{Contour detection and decoding}: The contour detection and decoding algorithm is applied to the binary image. This step is identical to Step (6) and (7) for thermal imaging. 
    
\end{enumerate}
Similar to the thermal imaging evaluations, we calculate the decoding accuracy as the performance for each condition. 

% Depth test.
\subsubsection{Information depth}
\hfill \\
We first evaluate the reading accuracy at different information depths. The information density (block size) and infill percentage are fixed to $5~mm$ and $10\%$ respectively. As clarified in Section~\ref{sec:method}, we use the surface-fill technique to fill the layers between the information model and the surface.

\begin{figure}[t]
    \centering
    \includegraphics[width=0.8\textwidth]{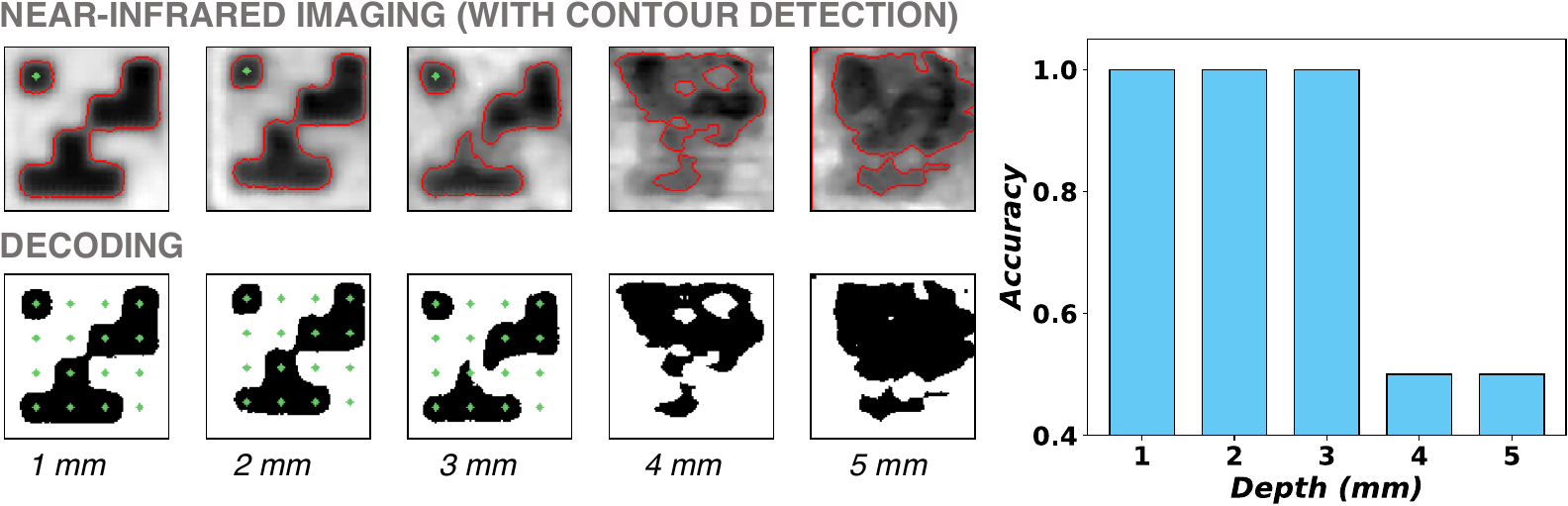}
    \caption{Near-infrared imaging and decoding results with different information depth.}
    \label{fig:eval_nirs_depth}
\end{figure}

The results are shown in Figure~\ref{fig:eval_nirs_depth}. It can be observed that the image quality decreases as the depth increases. For decoding, we succeeded in reading the binary data until depth $d=3~mm$. As scarce near-infrared lights penetrate deeper and reflect back ($d>3~mm$), the images become too blurry to read. It is worth noting that the information is visible with depth $d=1~mm$. Therefore, in practice, a depth $d\ge2~mm$ is suggested. We show more visibility tests in Section~\ref{subsec:visibility} and provide guidelines in Table~\ref{tab:design_space}.

% Size test.
\subsubsection{Information density}
\begin{figure}[t]
    \centering
    \includegraphics[width=0.8\textwidth]{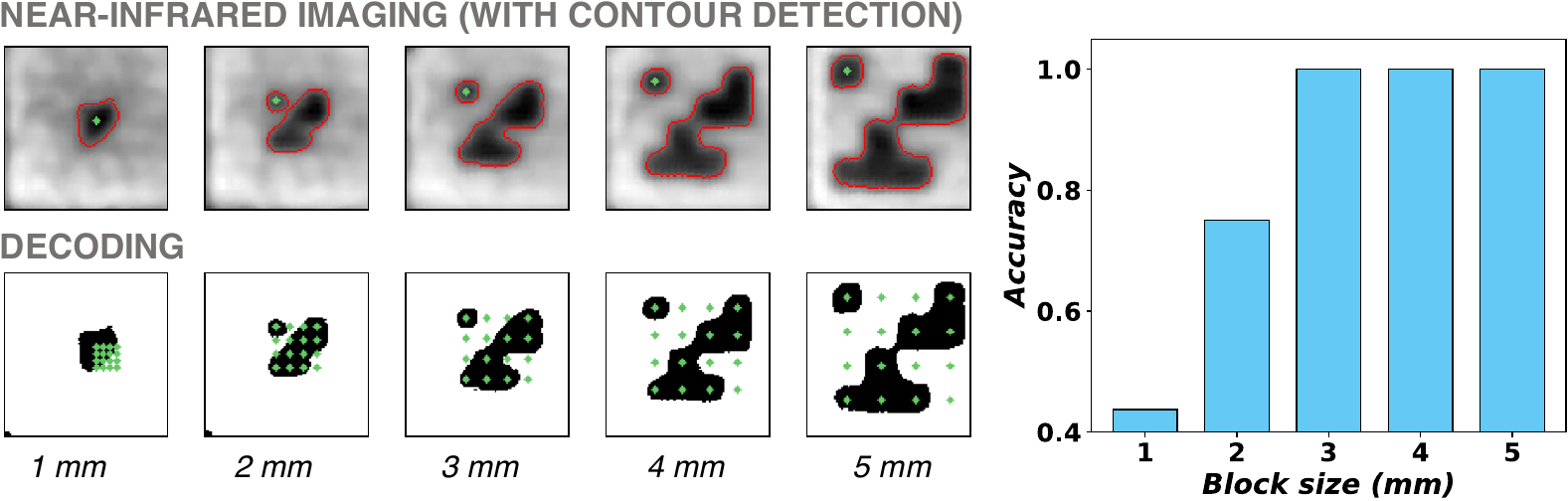}
    \caption{Near-infrared imaging and decoding results with different block sizes (information density).}
    \label{fig:eval_nirs_size}
\end{figure}

\hfill \\
Next, we test the information density by varying the block size of the information matrix. The information depth and infill percentage are fixed to $d=2~mm$ and $10\%$ respectively. The results are shown in Figure~\ref{fig:eval_nirs_size}. Our reading method can decode the data from block sizes as small as $X=3~mm$. Although the matrix itself can be successfully detected until $X=1~mm$, the decoding results are not perfect and the retrieved images are blurry. 

% Infill test.
\subsubsection{Infill percentage}
\begin{figure}[t]
    \centering
    \includegraphics[width=\textwidth]{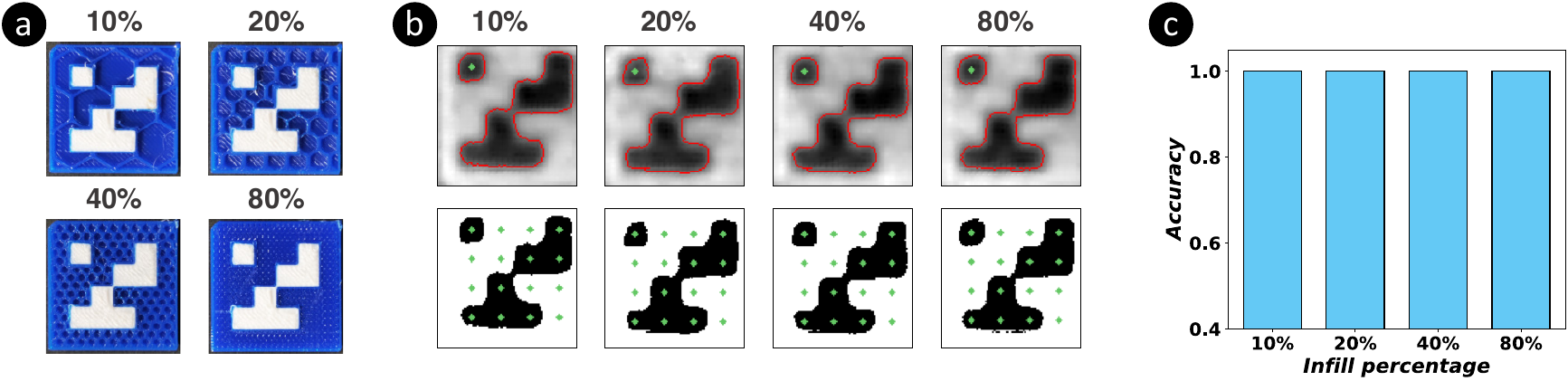}
    \caption{Near-infrared imaging with different infill percentages. \textbf{(a)} Cross-section for different infill percentages. Photos are transposed for illustrations. \textbf{(b)} Imaging (top row) and reading (bottom row) results. \textbf{(c)} Reading accuracy for the samples.}
    \label{fig:eval_nirs_infill}
\end{figure}

\hfill \\
Further, we evaluate the near-infrared imaging with different infill percentages. Aligned with the thermal imaging evaluation, we vary the infill percentages: $10\%$, $20\%$, $40\%$ and $80\%$. The information depth and block size are fixed at $d=2~mm$ and $X=5~mm$ respectively. The results are shown in Figure~\ref{fig:eval_nirs_infill}. The imaging results are similar, with perfect reading results for all conditions. This result confirms that the effect of the infill parameter is negligible for near-infrared imaging, as we use the aforementioned ``surface-fill'' technique for fabrications. 

% Color test.
\subsubsection{Color}
\begin{figure}[t]
    \centering
    \includegraphics[width=\textwidth]{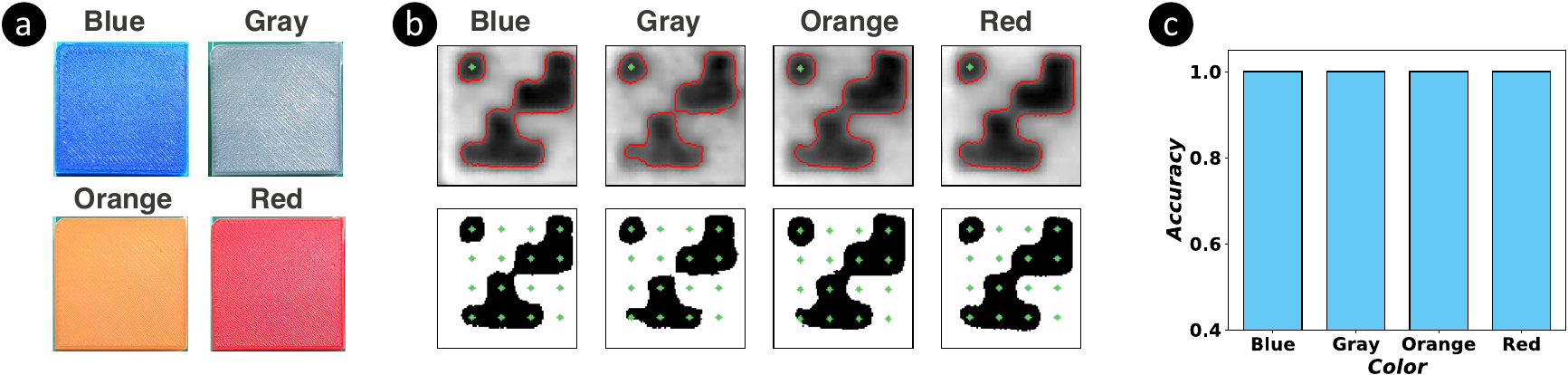}
    \caption{Near-infrared imaging and decoding results with different colors. \textbf{(a)} Samples printed using different colored materials. Photos are transposed for illustrations. \textbf{(b)} Imaging (top row) and reading (bottom row) results. \textbf{(c)} Reading accuracy for the samples.}
    \label{fig:eval_nirs_color}
\end{figure}

\hfill \\
In addition, we test the near-infrared imaging method using different colors. In particular, we vary the material color for the object (\ie the cube) while keeping the color for the information body as white. Five colors are chosen for testing: blue, gray, orange, red and black. We demonstrate the results in Figure~\ref{fig:eval_nirs_color}. For the non-black colors, the readings are all accurate. While for the black color, we cannot extract any information inside using near-infrared light. 

In principle, the black material absorbs the majority of light (both visible and infrared), and reflects very little. This characteristic makes the black material ideal for information embedding using thermal imaging, as we tested above. In contrast, the non-black materials are in fact translucent. Furthermore, near-infrared light can penetrate deeper than visible light, as we clarified in Section~\ref{sec:method}. This characteristic makes non-black colors more suitable for information embedding using the near-infrared scheme. 

% Show the visibility. 
\subsection{Visibility to the human eye}
\label{subsec:visibility}
Finally, we test the visibility of information embedding using different colors, as we aim to embed information in an unobtrusive manner. For the visibility tests, the following samples are prepared, considering to be practical scenarios:

\begin{enumerate}
    \item Thermal imaging samples: For thermal imaging, we fabricate the samples with information depth $d=1~mm$, with the ``surface-join'' fabrication technique. The same material is used for both the information model (\ie the matrix) and the object model (\ie the cube). 
    
    \item Near-infrared imaging samples: For near-infrared imaging, we vary the information depth in $d=1~mm$, $2~mm$ and $3~mm$. The information model is printed using PLA-white, while the object model is printed using different colors. 
\end{enumerate}

\input{tables/visibility_table}

For both thermal imaging samples and near-infrared imaging samples, we use the same color set above for the visibility test (\ie blue, gray, orange, red and black). In total, ($1$ thermal sample $+$ $3$ near-infrared samples) $\times$ $5$ colors $=$ $20$ samples are fabricated for the visibility tests. 

We classify the visibility into ``visible'', ``unobtrusive'' and ``invisible'', aligned with the literature~\cite{li2017aircode,delmotte2020blind}. For classification, two researchers first label the samples independently. For the controversial samples, a third researcher is included to decide the final labels independently. The ground-truth information is provided for reference. All samples are inspected in a well-illuminated office (illuminance $>350~lx$). The results are shown in Table~\ref{tab:visibility_result}. We run a Cohen's kappa test to measure the inter-rater reliability. The Cohen's kappa coefficient $\kappa = 0.76$, showed substantial agreement for the initial rating (3 out of 20 labels are controversial). For the samples printed by the ``surface-fill'' technique, we observe that the embedded information can be directly seen for depth $d=1~mm$. Whereas the embedded information is invisible for depth $d\ge2~mm$, except for the orange color (unobtrusive). For the ``surface-join'' case, only the black samples are invisible, while the blue and red samples are unobtrusive. 

% The table for design guideline. 
\input{tables/design_space}

\section{Discussion}
\subsection{Design guidelines}
\label{subsec:design_guideline}
% Might require user study. 
% Overall design space.

We have systematically explored multiple ways of embedding information in 3D objects, and provide a thorough evaluation of multiple parameters for fabrication and reading. Given our evaluation results, we now provide design guidelines for embedding information into 3D printed objects using an FDM 3D printer, as summarized in Table~\ref{tab:design_space}. We cover four design aspects for embedding information into a 3D printed object as follows:

\begin{enumerate}
    \setlength{\itemsep}{0.5em}
    \item \textit{Appearance: Whether the embedded information can be seen directly. } For designs with appearance-related constraints, we recommend to embed information at a depth of $d\ge2~mm$ with colored materials (refer to the example application shown in Figure~\ref{fig:demo_tools}). Alternatively, if the design can be fabricated using a black material, the information depth should be $d\le1~mm$ (refer to the example application in Figure~\ref{fig:demo_display}). 
    
    \item \textit{Functionality: Whether the printed objects are functional. } A typical functional print requires a high infill density. Since for thermal imaging the maximal infill percentage is $20\%$, we recommend using colored material incorporated with near-infrared imaging for embedding information into a functional printed object (refer to the example application shown in Figure~\ref{fig:demo_tools}).  
    
    \item \textit{Information density: Whether the embedded information requires high density. } For information density higher than (or pixel size smaller than) $5~mm~per~pixel$, we recommend the embedding technique with near-infrared imaging, using the ``surface-fill'' fabrication technique. In addition, a deep-learning based super-solution method can further increase the image quality for high-density information (refer to the example application shown in Figure~\ref{fig:demo_sign}).
    
    \item \textit{Reading speed: Whether an instant reading is required. } For embedded information that requires instant reading, we recommend embedding information incorporated with thermal imaging. Also, a thermal camera is more ubiquitous compared to a near-infrared device (\eg a mobile thermal camera~\cite{flir} or a smartphone with thermal camera~\cite{catphones}).
    
\end{enumerate}

% Limitations.
\subsection{Limitations and future work} 
\label{subsec:limitation}
We note several limitations of \textit{\systemname}, which need to be addressed in our future work and might further expand the design space and use cases.

\begin{itemize}[leftmargin=*]
    \setlength\itemsep{0.5em}
    \item We only tested \textit{\systemname} on an FDM printer. In principle, our method can also be applied to other 3D printing and digital fabrication techniques, such as SLA (Stereolithography) 3D printers and laser cutters that are also popular and feasible for practical use. In particular, incorporating multiple digital fabrication tools is considered a promising way to further expand the design space and accelerate design pipeline~\cite{beyer2015platener}. 
    
    \item The information density of \textit{\systemname} is not very high (up to $3~mm~per~pixel$). This limitation is mainly caused by the resolution of our imaging devices. As a complement, we utilized several computer vision algorithms to enhance the imaging quality. In the short term, the issue can be alleviated by tuning and adopting new computer vision methods that evolved rapidly in recent years~\cite{voulodimos2018deep}. In the long-term, as the hardware upgrades, the imaging quality improves, resulting in higher information density in the future. 
    
    \item Also, for the evaluation, we only included the binary case in a machine-readable way. For the texts and icons, the readability may vary among users. As we focus on the embedding technique in this work, it would be beneficial to run a user-study for testing the readability for different information types. 
    
    \item Our reading devices are currently not mobile. For thermal imaging, there are several mobile thermal cameras that are commercially available~\cite{flir}. We also observe the emergence of smartphones integrated with a thermal camera~\cite{catphones}. Mobile thermal cameras and smartphones with built-in thermal cameras should be tested in future work to determine the feasibility of \textit{\systemname} in everyday scenarios. Furthermore, for near-infrared imaging, we note that the already-popular depth camera for many recent smartphones using near-infrared wavelengths~\cite{grunnet2019intel} can potentially be used as a reading device for \textit{\systemname}. Future studies can focus on developing mobile apps utilizing those smartphone components. 
    
    \item For the information embedding scheme, our design is limited to a flat surface. In practice, many designs may not include a flat surface. Therefore, future work could explore a technique for embedding information under a curved surface. 
\end{itemize}

\section{Conclusion}
In this paper, we presented a technique to embed information invisible to the eye inside 3D printed objects. The information can be integrated and printed with the object directly, using off-the-shelf dual-head FDM 3D printers. For retrieving the information, we adopted and evaluated two methods, thermal imaging and near-infrared imaging, respectively. Based on the evaluation results, we proposed design guidelines of embedding information for different scenarios. Applications include interactive thermal displays, hidden board game tokens, tagging functional printed objects, and autographing non-fungible fabrication work, with more use cases enabled by expanding the design space of digital fabrications with our method and future work.

\bibliographystyle{ACM-Reference-Format}
\bibliography{main}

% Appendix
\appendix
\newpage
\onecolumn
% Reset counters.
\captionsetup[figure]{name=Appendix Fig.}
\captionsetup[table]{name=Appendix Table}
\setcounter{figure}{0}
\setcounter{table}{0}

\section*{Appendix}
\begin{figure}[h]
    \centering
    \includegraphics[width=\textwidth]{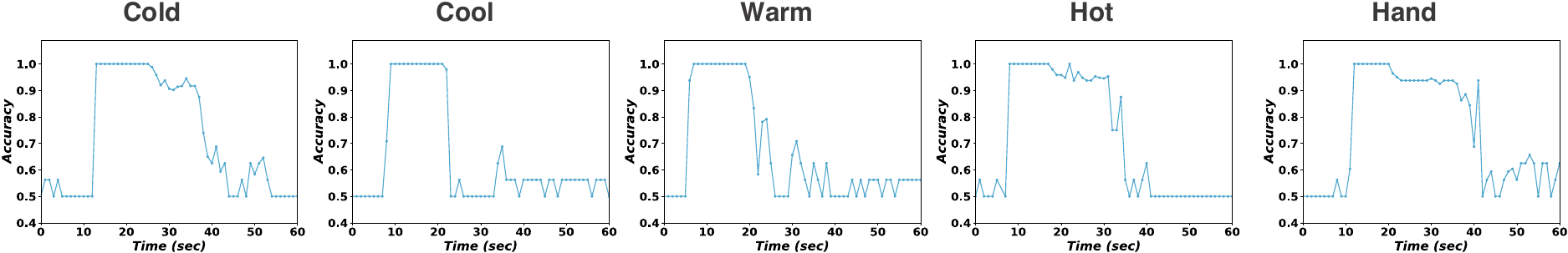}
    \caption{Thermal imaging evaluation results in different thermal transfer conditions.}
    \label{fig:eval_thermal_acc}
\end{figure}

\begin{figure}[h]
    \centering
    \includegraphics[width=\textwidth]{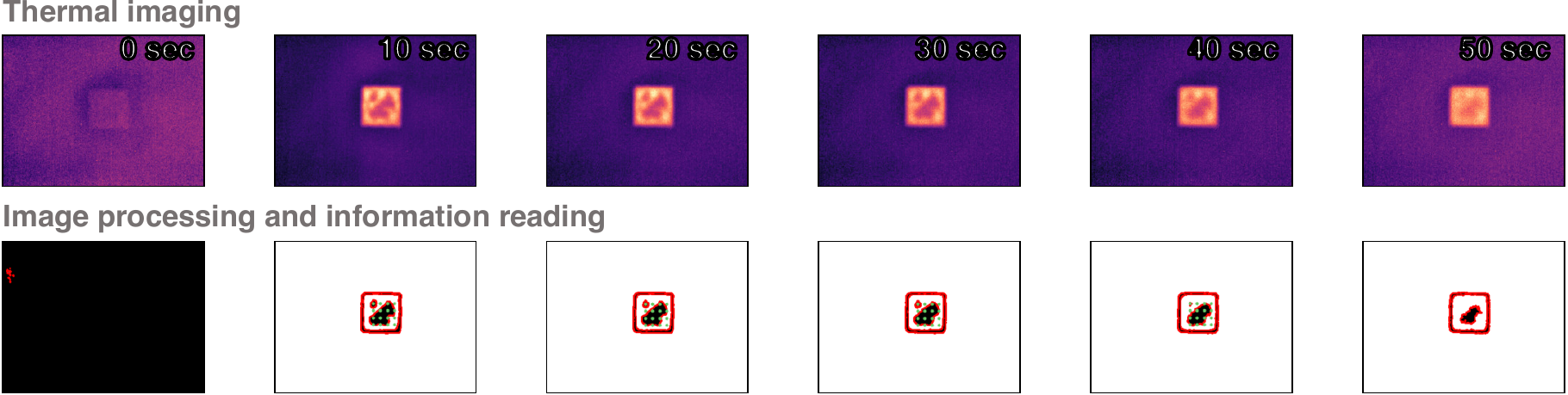}
    \caption{Illustration of thermal imaging results for information density = $4~mm$.}
    \label{fig:eval_thermal_w4}
\end{figure}

\begin{figure}[h]
    \centering
    \includegraphics[width=\textwidth]{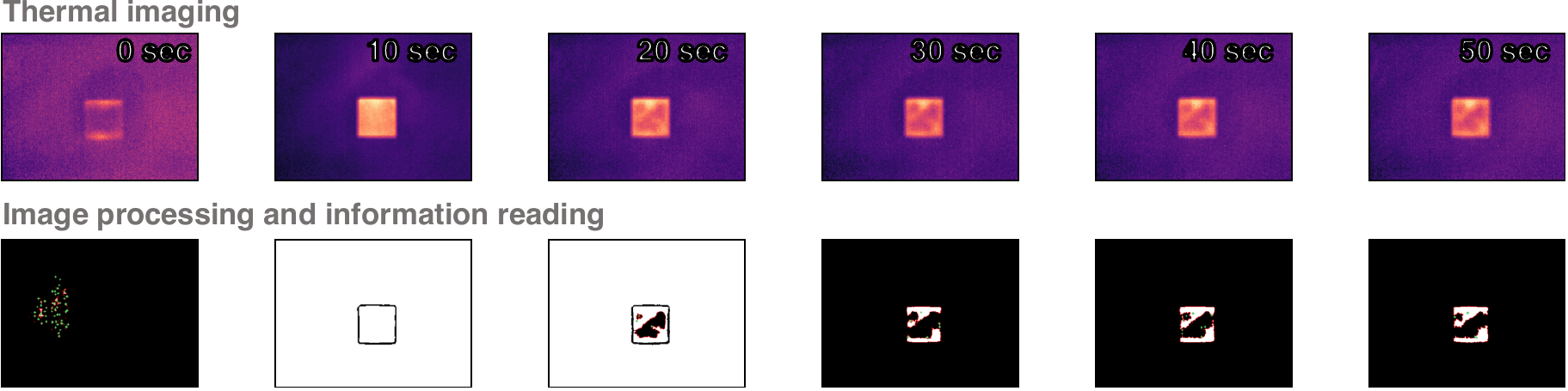}
    \caption{Illustration of thermal imaging results for information depth = $2~mm$.}
    \label{fig:eval_thermal_d2}
\end{figure}

\end{document}

%% file: tables/thermal_vs_nirs.tex
\begin{figure}[t]
\begin{minipage}[b]{0.8\textwidth}
    \centering
    \captionof{table}{Comparison between thermal (mid- to far-infrared) and near-infrared imaging methods}
    \vspace{-0.3em}
    \noindent
    \resizebox{1.0\textwidth}{!}{%
    \begin{tabular}{l l l l l r}
        \toprule
        \bf Imaging Method & \bf Reading Speed & \bf Information Density & \bf Depth & \bf Infill Density & \bf \makecell[r]{Color} \\ \toprule
        \it \makecell[l]{Thermal\\ (Mid- to far-infrared)} & Instant & $\ge5~mm$ per pixel & $\le1~mm$ & $\le20\%$ & \makecell[r]{Any color} \\ \midrule
        \it Near-infrared & Tens of minutes & $\ge3~mm$ per pixel & $\le3~mm$ & Any & \makecell[r]{Non-black (outside)\\ + white (inside)} \\ \toprule
    \end{tabular}}
    \label{tab:thermal_vs_nirs}
\end{minipage}
% \hspace{1em}
% Ground-truth
\begin{minipage}[b]{0.18\textwidth}
    \centering
    \includegraphics[width=0.3\textwidth]{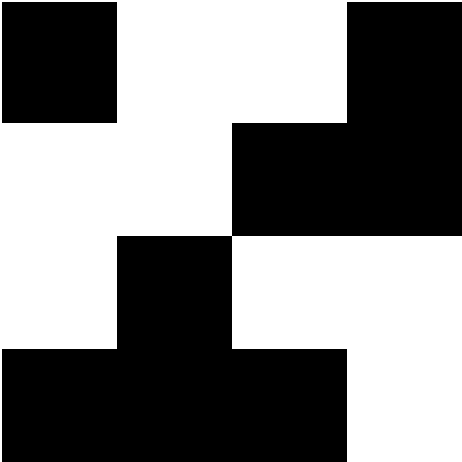}
    \captionof{figure}{Ground truth}
    \label{fig:sample_groundtruth}
\end{minipage}
\end{figure}

%% file: tables/visibility_table.tex
\begin{table}[t]
\centering
\caption{Visibility test results for different samples.}
\vspace{-1em}
\newcolumntype{Y}{>{\centering\arraybackslash}X}

\begin{tabularx}{0.83\textwidth}{XYYYY}
    \toprule
     & \multicolumn{3}{c}{\textbf{Surface-fill} (near-infrared imaging)} &  \\ \cline{2-4}
    \multirow{-2}{*}{\bf Color} & \it d=1 mm & \it d=2 mm & \it d=3 mm & \multirow{-2}{*}{\makecell{\bf Surface-join \\ (thermal imaging)}} \\ \toprule
    Blue & \cellcolor[HTML]{F0C6C6}Visible & \cellcolor[HTML]{BBFFBA}Invisible & \cellcolor[HTML]{BBFFBA}Invisible & \cellcolor[HTML]{FFCC67}Unobtrusive \\
    Red & \cellcolor[HTML]{F0C6C6}Visible & \cellcolor[HTML]{BBFFBA}Invisible & \cellcolor[HTML]{BBFFBA}Invisible & \cellcolor[HTML]{FFCC67}Unobtrusive \\
    Orange & \cellcolor[HTML]{F0C6C6}Visible & \cellcolor[HTML]{FFCC67}Unobtrusive & \cellcolor[HTML]{BBFFBA}Invisible & \cellcolor[HTML]{F0C6C6}Visible \\
    Gray & \cellcolor[HTML]{F0C6C6}Visible & \cellcolor[HTML]{BBFFBA}Invisible & \cellcolor[HTML]{BBFFBA}Invisible & \cellcolor[HTML]{F0C6C6}Visible \\
    Black & \cellcolor[HTML]{BBFFBA}Invisible & \cellcolor[HTML]{BBFFBA}Invisible & \cellcolor[HTML]{BBFFBA}Invisible & \cellcolor[HTML]{BBFFBA}Invisible \\ \toprule
    \end{tabularx}
\label{tab:visibility_result}

\end{table}

%% file: tables/design_space.tex
% Please add the following required packages to your document preamble:
% \usepackage{multirow}
\begin{table}[t]
    \centering
    \renewcommand{\arraystretch}{1.5}
    \setstretch{1.2}
    \caption{Guidelines of information embedding into 3D prints}
    \vspace{-0.5em}
    
    \resizebox{\linewidth}{!}{%
    \begin{tabular}{llcccp{0.31\linewidth}}
        \toprule
        \multirow{2}{*}{\textbf{Design aspect}} & \multirow{2}{*}{\textbf{Design target}} & \multicolumn{3}{c}{\textbf{Design space}} & \multirow{2}{*}{\textbf{Examples / use cases}} \\ \cline{3-5}
         &  & \textit{Color} & \textit{Depth (d)} & \textit{Imaging} &  \\ \toprule
        \multirow{4}{*}{\vspace{-2em} Appearance} & Visible & Non-black & $\le1~mm$ & - & \makecell[l]{Generic tagged objects with \\ visible under-surface information} \\ \cline{2-6} 
         & Unobtrusive & Non-black & $1~mm<d\le2~mm$ & Near-infrared & \makecell[l]{Tagged artworks, models,\\designs requiring aesthetics} \\ \cline{2-6}
         & \multirow{2}{*}{\vspace{-0.8em} Invisible} & Non-black & $2~mm\le d \le3~mm$ & Near-infrared & \makecell[l]{Tagged artworks, models,\\designs requiring aesthetics} \\ \cline{3-6} 
         &  & Black & $d\le1~mm$ & Thermal & \makecell[l]{Tangible \& invisible tokens,\\ dynamic displays} \\ \hline
        \multirow{3}{*}{\vspace{-2em} Functionality} & \multirow{2}{*}{Non-functional} & Non-black & $d\le3~mm$ & \textit{Either}\footnotemark[1] & \multirow{2}{*}{\makecell[l]{Generic tagged objects\\ (standard prints)}} \\ \cline{3-5}
         &  & Black & $d\le1~mm$ & Thermal &  \\ \cline{2-6} 
         & Functional & Non-black & $d\le3~mm$ & Near-infrared & \makecell[l]{Tagged tools, \\long-term use components} \\ \hline
        \multirow{3}{*}{\makecell[l]{\\Information density\\($X,~mm~per~pixel$)}} & \multirow{2}{*}{$X\ge5$} & Non-black & $d\le3~mm$ & Near-infrared & \multirow{2}{*}{\makecell[l]{Objects embedded with large \\ icons, texts or low-density data}} \\ \cline{3-5}
         &  & Black & $d\le1~mm$ & Thermal &  \\ \cline{2-6} 
         & $3\le X \le5$ & Non-black & $d\le3~mm$ & Near-infrared & \makecell[l]{Non-fungible autographed prints,\\objects embedded with small \\icons, texts or high-density data.} \\ \hline
        \multirow{2}{*}{\vspace{-2em} Reading speed} & Instant & Black & $d\le1~mm$ & Thermal & \makecell[l]{Board game token, dynamic display,\\ tags requiring instant reading} \\ \cline{2-6} 
         & Not instant & Non-black & $d\le3~mm$ & \textit{Either}\footnotemark[1] & \makecell[l]{Generic objects tagged under the \\ surface} \\ \toprule
    \end{tabular}}
    
    \label{tab:design_space}
    {\raggedright \small \footnotemark[1] Refer to Table~\ref{tab:thermal_vs_nirs} for more details. \par}
    \vspace{-1.5em}
\end{table}